\documentclass[showpacs,aps,graphicx,twocolumn]{revtex4}

\usepackage{amssymb}
\usepackage{graphicx}

\begin{document}

\title{Fault tolerant quantum key distribution based on  quantum dense coding with collective noise\footnote{Published in
Int. J. Quant. Inform. 7 (8), 1479-1489 (2009).}}
\author{ Xi-Han Li$^{a,b}$, Bao-Kui Zhao$^{a,b}$, Yu-Bo Sheng$^{a,b}$,
 Fu-Guo Deng$^{c}$\footnote{E-mail address: fgdeng@bnu.edu.cn.} and  Hong-Yu Zhou$^{a,b}$}
\address{$^a$ Key Laboratory of Beam Technology and Material Modification of Ministry of Education,
Beijing Normal University,
Beijing 100875,  China\\
$^b$ College of Nuclear Science and Technology, Beijing Normal
University,
Beijing 100875, China\\
$^c$ Department of Physics, Beijing Normal University, Beijing
100875, China}
\date{\today}

\begin{abstract}
We present two robust quantum key distribution protocols against two
kinds of collective noise, following some ideas in  quantum dense
coding. Three-qubit entangled states are used as quantum information
carriers, two of which forming the logical qubit which is invariant
with a special type of collective noise. The information is encoded
on logical qubits with four unitary operations, which can be read
out faithfully with Bell-state analysis on two physical qubits and a
single-photon measurement on the other physical qubit, not
three-photon joint measurements. Two bits of information are
exchanged faithfully and securely by transmitting two physical
qubits through a noisy channel. When the losses in the noisy channel
is low, these protocols can be used to transmit a secret message
directly in principle.
\end{abstract}

 \maketitle

\section{introduction}

Quantum dense coding and quantum teleportation are two typical
examples in quantum information exploiting striking nonclassical
properties of entangled states to perform otherwise impossible
tasks. Quantum dense coding, which was first proposed in 1992
\cite{dc}, enables the communication of two bits of classical
information with the transmission of a qubit. This feature makes it
have a good application in secure quantum communication
\cite{high,caipra}. Quantum dense coding was demonstrated by using
photon pairs entangled in polarization in 1996 \cite{dce} and
subsequently was investigated experimentally by using some other
quantum systems, such as continuous variables \cite{continuous},
nuclear magnetic resonance \cite{nuclear} and atomic qubits
\cite{atomic}. The theoretical researches have been broadened to
high-dimension quantum states \cite{high}, more than two entangled
qubits \cite{multi} and hyperentangled quantum states \cite{hyper}.

In quantum dense coding, the Bell-state analysis (BSA) is one of
the key important steps. Although the four Bell states are
orthogonal to each other and is deemed to be discriminated
deterministically in principle, BSA is always a technical
difficulty, which yielding low efficiencies and low discrimination
fidelities. Using only the polarization degree of freedom of
photons, three Bell states can at best be discriminated with
linear optics, which reduces the attainable channel capacity of
quantum dense coding from 2 bits to log$_2$3 $\approx$ 1.585 bits
\cite{dce}. Recently, cross-Kerr nonlinearity was used to
implement two-qubit controlled-NOT gate  \cite{kerr}, whose
character can also be used to solve the BSA problem. Furthermore,
hyperentanglement (i.e., entanglement in more than one degree of
freedom (DOF)) enables a full BSA in polarization DOF with linear
optics \cite{hyper,momentum}. In particular, Schuck  et al.
\cite{complete} proposed a complete deterministic linear-optics
BSA by employing the intrinsic time-energy correlation of photon
pairs generated with high temporal definition in 2006. If we can
control the time of arrival of these two photons at beam splitter
(BS), i.e, make them arrive at the BS simultaneously, these four
Bell states can be discriminated faithfully in principle and the
experiment gives its success probability between 81\%-89\%.

Although there are many kinds of physical systems which can be
selected by quantum communication, photon is the most popular one,
relying on its good feature in preparation and manipulation.
However, during the transmission, the polarization DOF of photons is
incident to be influenced by the thermal fluctuation, vibration and
the imperfection of the fiber, which we call them noise in total. At
present, we always suppose the noise in a quantum channel is a
collective one \cite{collective}. That is, the fluctuation of noise
is slow in time. If several photons transmit through the noisy
channel simultaneously or they are close to each other, the
alternation arising from the noise on each qubit is identical. With
this kind of noise, several methods have been proposed to cancel or
reduce the noise effect, such as entanglement purification
\cite{ep}, quantum error correct code (QECC) \cite{qecc},
single-photon error rejection \cite{er} and decoherence-free
subspace (DFS) \cite{dfs1,qkd}. QECC encodes one logical bit into
several physical qubits according to the kind of noise, and the user
measures the stabilizer codes to detect the errors and then correct
them. Compared with QECC, single-photon error rejection schemes
require less quantum resource, although they  succeed
probabilistically. Entanglement purification is a method to distill
a maximally entangled states from some less-entanglement states by
sacrificing several samples affected by the noise. The influence of
noise can only be reduced in entanglement purification and infinite
steps are needed to get a perfect maximally entangled state. DFS can
be composed of several qubits which suffer from the same noise and
compensate the effect of noise, and then it posses the invariability
against the noise. This feature can be used to encode messages in
quantum communication over a noisy channel. Boileau  et al.
\cite{qkd} proposed a quantum key distribution (QKD) scheme with a
collective random unitary noise by using linear combinations of two
singlet states $\vert \psi^- \rangle$.

In this paper, we present two fault-tolerant quantum key
distribution schemes against two different kinds of collective
noise, the collective-dephasing noise and the collective-rotation
one, following some ideas in quantum dense coding. Three-qubit
entangled states are utilized as quantum information carriers. Each
logical qubit transmitted is composed of two physical qubits and is
invariant during the transmission. The message is encoded on each
logical qubit transmitted with one of four unitary operations and
can be read out with a Bell-state analysis and a single-photon
measurement on each three-qubit quantum system, not a three-qubit
joint measurement (i.e., a Greenberger-Horne-Zeilinger (GHZ) state
measurement). Two bits of information are exchanged faithfully and
securely by transmitting two physical qubits through a noisy
channel. When the losses in the noisy channel is low, our quantum
cryptography schemes can, in principle, be used to transmit a secret
message directly.

\section{Fault tolerant quantum key distribution based on quantum dense coding  with collective noise}

\subsection{review of  quantum cryptography based on quantum dense coding in an ideal condition}

We review the basic principle of quantum cryptography based on
quantum dense coding  in an ideal condition first
\cite{high,caipra}. Suppose the receiver Alice wants to share a set
of key with the sender Bob. She prepares a sequence of
Einstein-Podolsky-Rosen (EPR) pairs in which each is randomly in one
of four Bell states
\begin{eqnarray}
\vert \phi^\pm \rangle_{AB} =\frac{1}{\sqrt{2}}(\vert 00 \rangle
\pm
\vert 11 \rangle)_{AB},\\
\vert \psi^\pm \rangle_{AB}=\frac{1}{\sqrt{2}}(\vert 01 \rangle
\pm \vert 10 \rangle)_{AB},
\end{eqnarray}
where the subscripts $A$ and $B$ represent two entangled photons.
$\vert 0 \rangle$ and $\vert 1 \rangle$ are the two eigenvectors of
$\sigma_z$. Alice picks up photon $B$ in each pair to form the
sequence $S_B$ and sends it to Bob who performs one of four unitary
operations on each photon in  the sequence $S_B$ randomly. These
four operations are
\begin{eqnarray}
U_{00}&=&U_I=I=\vert 0 \rangle \langle 0 \vert + \vert 1 \rangle
\langle 1 \vert,\nonumber\\
U_{01}&=&U_z=\sigma_z=\vert 0 \rangle \langle 0 \vert - \vert
1\rangle \langle 1 \vert,\nonumber\\
U_{10}&=&U_x=\sigma_x=\vert 1 \rangle \langle 0 \vert + \vert
0\rangle \langle 1 \vert,\nonumber\\
U_{11}&=&U_y=-i\sigma_y=\vert 1 \rangle \langle 0 \vert - \vert
0\rangle \langle 1 \vert.
\end{eqnarray}
The subscripts 00, 01, 10 and 11 are the two bits of information
represented by the corresponding operations. $\sigma_x$, $\sigma_y$,
$\sigma_z$ are Pauli operators. These four unitary operations will
change the four Bell states from one to another. For example,
operators $U_{00}$, $U_{01}$, $U_{10}$ and $U_{11}$ can transform
the state $\vert \phi^+ \rangle$ into $\vert \phi^+ \rangle$, $\vert
\phi^- \rangle$, $\vert \psi^+ \rangle$, and $\vert \psi^- \rangle$,
 respectively. After the operations,  Bob transmits the sequence $S_B$
 back to Alice who knows the original states and performs
two-photon Bell-state measurements on the photon pairs. In this way,
Alice can get the information of operations Bob performed and she
can share a sequence of secret key with Bob. As two bits of key can
be shared by means of only one photon being transmitted, this
quantum information process is named as quantum dense coding.

When there is noise in the transmission of photons, we can use a
similar entangled state to implement quantum dense coding, which
can be written in a general form
\begin{eqnarray}
\vert \Phi^\pm \rangle_{AB}= \frac{1}{\sqrt{2}}(\vert 0 \rangle_A
\vert 0_L \rangle_B \pm \vert 1 \rangle_A \vert 1_L \rangle_B),\nonumber\\
\vert \Psi^\pm \rangle_{AB}= \frac{1}{\sqrt{2}}(\vert 0 \rangle_A
\vert 1_L \rangle_B \pm \vert 1 \rangle_A \vert 0_L \rangle_B).
\end{eqnarray}
Here the subscript $L$ denotes the logical qubit, which can be made
of several physical qubits experiencing the same noise and then
compensating the noise effect due to the peculiar relationship
between them. The logical qubits are invariant with the noise and
are constructed according to the form of noise.

It is well known that quantum dense coding can be used to improve
the capacity of secure quantum communication \cite{high,caipra}. The
security of quantum communication is its most important character,
compared with classical communication. If quantum dense coding
cannot be used for secure quantum communication, it will loss the
significance as a classical signal is far easier to be prepared and
robust against the channel noise. For instance, in a classical way,
Alice and Bob can communicate easily with the two states $|0\rangle$
and $|1\rangle$ against a collective-dephasing noise shown in
Eq.(\ref{dephasing}). Also they can communicate by using the two
states $\vert +y\rangle=\frac{1}{\sqrt{2}}(\vert 0\rangle + i\vert
1\rangle)$ and $\vert -y\rangle=\frac{1}{\sqrt{2}}(\vert 0\rangle -
i\vert 1\rangle)$ to avoid the effect of the collective-rotation
noise shown in Eq.(\ref{rotation}). However, a secure quantum
communication resorts to at least two nonorthogonal states
\cite{bb84,b92,Ekert91,BBM92,rmp,longqkd,CORE,BidQKD,lixhqkd}. That
is, it requires that the superposition of two eigenvectors is also
free to the collective noise. The two classical states $\{\vert
0\rangle, \vert 1\rangle\}$ are immune to a collective-dephasing
noise, but their superposition is not. So do the two states $\{\vert
+y\rangle, \vert -y\rangle\}$ to a collective-rotation noise. We
will introduce two methods to construct a logical qubit against two
kinds of collective noise, which will retain the security of quantum
cryptography simultaneously. We discuss them in detail below.

\subsection{Quantum cryptography based on quantum dense coding against a collective-dephasing noise}

A collective-dephasing noise can be described as
\begin{eqnarray}
U_{dp}\vert 0 \rangle = \vert 0 \rangle, \;\;\;\; U_{dp}\vert 1
\rangle= e^{i\phi}\vert 1 \rangle,\label{dephasing}
\end{eqnarray}
where $\phi$ is the noise parameter and it fluctuates with time.
In general, a logical qubit encoded into two physical qubits with
antiparallel parity is immune to this kind of noise as these two
logical qubits acquire the same phase factor $e^{i\phi}$
\begin{eqnarray}
\vert 0_L \rangle =\vert 01\rangle,  \;\;\;\; \vert 1_L \rangle=
\vert 10 \rangle.
\end{eqnarray}
In this way,  the four entangled logical states used for quantum
cryptography based on quantum dense coding against a
collective-dephasing noise can be made up of four three-qubit
entangled states as
\begin{eqnarray}
\vert \Phi_{dp}^\pm \rangle_{AB_1B_2}= \frac{1}{\sqrt{2}}(\vert 0
\rangle_{A}
\vert 01 \rangle_{B_1B_2} \pm \vert 1 \rangle_{A} \vert 10 \rangle_{B_1B_2}),\nonumber\\
\vert \Psi_{dp}^\pm \rangle_{AB_1B_2}= \frac{1}{\sqrt{2}}(\vert 0
\rangle_{A} \vert 10 \rangle_{B_1B_2} \pm \vert 1 \rangle_{A}
\vert 01 \rangle_{B_1B_2}) \label{1}.
\end{eqnarray}
The logical qubit $B$ is composed of two physical qubits $B_1$ and
$B_2$.

Our quantum cryptography scheme based on quantum dense coding
against a collective-dephasing noise works as follows.

(1) Similar to Refs. \cite{high,longqkd}, Alice prepares a sequence
of quantum systems in the entangled state $\vert \Phi^+
\rangle_{AB}= \frac{1}{\sqrt{2}}(\vert 0 \rangle \vert 0_L \rangle +
\vert 1 \rangle \vert 1_L \rangle)_{AB}=\frac{1}{\sqrt{2}}(\vert 0
\rangle_{A} \vert 01 \rangle_{B_1B_2} + \vert 1 \rangle_{A} \vert 10
\rangle_{B_1B_2})$. Alice divides the quantum systems into two
photon sequences, $S_A$ and $S_B$. $S_A$ is composed of all the
logical qubits $A$ in the quantum systems and $S_B$ is made up of
the logical qubits $B$, which is composed of two physical qubits
$B_1$ and $B_2$.

(2) Alice sends the logical qubit sequence $S_B$ to Bob and keeps
the sequence $S_A$.

(3) After receiving the sequence $S_B$, Bob chooses randomly a
subset of logical qubits as the samples for error rate analysis and
the other logical qubits make up of the message sequence $S_{BM}$
for generating the private key. Bob measures each logical qubit in
the samples by choosing randomly one of the two bases $\sigma_z^B
\equiv \sigma_z^{B_1}\otimes \sigma_z^{B_2}$ and $\sigma_x^B \equiv
\sigma_x^{B_1}\otimes \sigma_x^{B_2}$. Also he tells Alice which
logical qubits are chosen as samples. Under these two bases, the
original entangled state can be written as
\begin{eqnarray}
\vert \Phi_{dp}^+ \rangle_{AB_1B_2} &=& \frac{1}{\sqrt{2}}(\vert 0
\rangle_{A} \vert 01 \rangle_{B_1B_2} + \vert 1 \rangle_{A} \vert
10 \rangle_{B_1B_2})\nonumber\\
&=& \frac{1}{2}[\vert + \rangle_{A} (\vert ++ \rangle - \vert --
\rangle)_{B_1B_2} \nonumber\\
&& \; + \vert - \rangle_{A} (\vert -+ \rangle - \vert +-
\rangle)_{B_1B_2}].
\end{eqnarray}
That is, the outcomes obtained by Alice and Bob are correlated if
they choose the same bases for their logical qubits. However, the
vicious action done by  Eve will destroy this correlation and be
detected, same as those in Refs.\cite{BBM92,twostep}. Alice and Bob
can exploit the correlation to check the security of the
transmission of the logical qubits run from Alice to Bob.

(4) For the message sequence $S_{BM}$, Bob operates each logical
qubit with one of the four unitary operations,
$\{\Omega_{00},\Omega_{01},\Omega_{10},\Omega_{11}\}$ which is made
up of two unitary operations on two physical qubits, i.e.,
\begin{eqnarray}
\Omega_{00}&=&\Omega_I=I_1\otimes I_2,\nonumber\\
\Omega_{01}&=&\Omega_z=U_{z1}\otimes I_2,\nonumber\\
\Omega_{10}&=&\Omega_x=U_{x1} \otimes U_{x2},\nonumber\\
\Omega_{11}&=&\Omega_y=U_{y1} \otimes U_{x2}.
\end{eqnarray}
The subscripts of $\Omega$ denotes the code $\{00, 01, 10, 11\}$,
and the subscripts 1 and 2 represent the unitary operations
performing on the photons $B_1$ and $B_2$, respectively. The
relation between the initial state, final state and the operator is
shown in Table I. These four operations will not drive the states
prepared by Alice out of the subspace $S=\{\vert \Phi_{dp}^\pm
\rangle_{AB_1B_2}, \vert \Psi_{dp}^\pm \rangle_{AB_1B_2}\}$. In
other words, the states operated by Bob are also antinoise when they
are transmitted back to Alice through a collective-dephasing noise
channel. After the unitary operation, Bob sends the sequence
$S_{BM}$ back to Alice.

\begin{center}
\begin{table}[!ht]
\label{table1} \caption{The relation between the initial state
$\vert \psi \rangle_i$, the final state $\vert \psi \rangle_f$ and
the unitary operation.}
\begin{tabular}{c|cccc}\hline
& $\vert \Phi^+_{dp} \rangle_f$ & $\vert \Phi^-_{dp} \rangle_f$ &
$\vert \Psi^+_{dp} \rangle_f$ & $\vert \Psi^-_{dp} \rangle_f$
\\\hline $\vert \Phi^+_{dp} \rangle_i$ & $\Omega_{00}$ & $\Omega_{01}$ &
$\Omega_{10}$ &
$\Omega_{11}$\\
$\vert \Phi^-_{dp} \rangle_i$ & $\Omega_{01}$ & $\Omega_{00}$ &
$\Omega_{11}$ &
$\Omega_{10}$\\
$\vert \Psi^+_{dp} \rangle_i$ & $\Omega_{10}$ & $\Omega_{11}$ &
$\Omega_{00}$ &
$\Omega_{01}$\\
$\vert \Psi^-_{dp} \rangle_i$ & $\Omega_{11}$ & $\Omega_{10}$ &
$\Omega_{01}$ &
$\Omega_{00}$\\
\hline
\end{tabular}
\end{table}
\end{center}

(5) Alice picks up each entangled quantum system from the two photon
sequences $S_{AM}$ and $S_{BM}$. Here $S_{AM}$ is made up of the
remaining logical qubits in $S_{A}$ after the first sampling for
analyzing error rate of the first transmission from Alice to Bob.
Alice performs a Bell-state measurement on the two photons $AB_1$
and a single-photon measurement on the photon $B_2$ with
$X=\sigma_x$ basis for distinguishing the four GHZ states $\vert
\Phi_{dp}^\pm \rangle_{AB_1B_2}$ and $\vert \Psi_{dp}^\pm
\rangle_{AB_1B_2}$. These four states to be discriminated can be
written as
\begin{eqnarray}
\vert \Phi_{dp}^\pm \rangle_{AB_1B_2}= \frac{1}{\sqrt{2}} (\vert
\phi^\pm \rangle_{AB_1}\otimes \vert + \rangle_{B_2}-\vert \phi^\mp
\rangle_{AB_1}
\otimes\vert - \rangle_{B_2}),\nonumber\\
\vert \Psi_{dp}^\pm \rangle_{AB_1B_2}= \frac{1}{\sqrt{2}} (\vert
\psi^\pm \rangle_{AB_1}\otimes \vert + \rangle_{B_2}+\vert \psi^\mp
\rangle_{AB_1}\otimes\vert - \rangle_{B_2}).\nonumber\\
\end{eqnarray}
Here $\vert \pm \rangle=(1/\sqrt{2})(\vert 0 \rangle \pm \vert 1
\rangle)$ are two eigenvectors of Pauli operator $\sigma_x$. From
this expression, one can see that with these two measurement
outcomes Alice can distinguish these four three-photon GHZ states
deterministically and read out the operations done by Bob on the
logical qubits $B$ in principle.

(6) For checking the security of the second transmission from Bob to
Alice, Alice chooses randomly a subset of the outcomes of the
measurements on three-qubit states operated by Bob as the samples
for error rate analysis, similar to Ref.\cite{high}. She requires
Bob tell her his operations on these samples and then analyzes the
security of the second transmission by herself. If the error rate is
very low, she tells Bob that their quantum communication is secure
in principle; otherwise, they will discard their outcomes and repeat
their quantum communication from the beginning.

(7) If their quantum communication is secure, Alice and Bob exploit
error correction and privacy amplification techniques to distil a
private key, same as Ref. \cite{rmp}.

As the logical qubits are always in the maximal mixture of $\vert
0_L \rangle$ and $\vert 1_L \rangle$ during the transmission, the
intercept-resending attack strategy by Eve will not work. The secret
message are encoded on logical qubits with four unitary operations,
which may be eavesdropped by inserting spy photons \cite{spy}, in
particular by inserting delay photons as the invisible photons with
a different wavelength compared to the legitimate photons can be
filtrated out by a special filter \cite{spy} before Bob performs his
operations. In order to guarantee the security, Bob has to select
parts of samples to check the multiphoton $(n>2)$ rate as the attack
with spy photons will increase the number of photons in each quantum
signal. In detail, Bob can choose another subset of logical qubits
$B$ and analyzes its multiphoton rate with photon number splitters
(PNSs). That is, for each logical qubit $B$, Bob uses two PNSs to
split each physical qubit ($B_1$ and $B_2$) and then measures the
number of the photons in each legitimate physical qubit. If the
multiphoton rate is unreasonably high, Bob terminates the
transmission and they repeats the communication from the beginning,
same as Ref.\cite{spy}. Certainly, the technique of photon number
splitters is not mature in the application of quantum communication.
At present, Bob can use beam splitters instead of PNSs to complete
the task of analyzing the multiphoton rate, same as Ref.\cite{spy}.

Without Trojan horse attack with spy photons, the transmission of
the sequence $S_B$ from Alice to Bob is completely same as that in
Bennett-Brassard-Mermin 1992 (BBM92) quantum key distribution (QKD)
protocol \cite{BBM92} which has been proven secure
\cite{prove1,prove2}. Thus, there is no problem in the security of
the transmission of the sequence $S_B$ from Alice to Bob in
principle. On the other hand, the reduced matrix of each logical
qubit $B$ from Bob to Alice is $\rho_{B}=\frac{1}{2}\left(
\begin{array}{cc}
1 & 0 \\
0 & 1
\end{array}
\right)$, which means that Eve cannot get useful information if she
only eavesdrops the transmission from Bob to Alice in principle.
That is to say, our quantum cryptography based on fault tolerant
quantum dense coding against a collective-dephasing noise is secure
in principle.

\subsection{Quantum cryptography based on quantum dense coding against a collective-dephasing noise}

Another kind of noise model called a collective-rotation noise
operates as
\begin{eqnarray}
U_r \vert 0 \rangle &=& \textrm{cos} \theta \vert 0 \rangle +
\textrm{sin} \theta
\vert 1\rangle,\nonumber\\
U_r \vert 1 \rangle &=& -\textrm{sin} \theta \vert 0 \rangle +
\textrm{cos} \theta \vert 1 \rangle.\label{rotation}
\end{eqnarray}
The parameter $\theta$ depends on the noise and fluctuates with
time. With such a type of collective noise, $\vert \phi^+ \rangle$
and $\vert \psi^-\rangle$ are invariant. Logical qubits can be
chosen as
\begin{eqnarray}
\vert 0_L \rangle= \vert \phi^+ \rangle,  \;\;\;\;  \vert 1_L
\rangle =\vert \psi^- \rangle.
\end{eqnarray}
Then the states for quantum cryptography based on quantum dense
coding can be chosen as
\begin{eqnarray}
\vert \Phi_{r}^\pm \rangle_{AB_1B_2}= \frac{1}{\sqrt{2}}(\vert 0
\rangle_{A}
\vert \phi^+ \rangle_{B_1B_2} \pm \vert 1 \rangle_{A} \vert \psi^- \rangle_{B_1B_2}),\nonumber\\
\vert \Psi_{r}^\pm \rangle_{AB_1B_2}= \frac{1}{\sqrt{2}}(\vert 0
\rangle_{A} \vert \psi^- \rangle_{B_1B_2} \pm \vert 1 \rangle_{A}
\vert \phi^+ \rangle_{B_1B_2}). \label{2}
\end{eqnarray}
Also, Alice divides the quantum systems into two photon sequences,
$S_A$ and $S_B$. She sends the logical qubit sequence $S_B$ to Bob
and keeps the sequence $S_A$. After receiving the sequence $S_B$,
Bob chooses randomly a subset of logical qubits as the samples for
error rate analysis and the other logical qubits make up of the
message sequence $S_{BM}$ for generating the private key. Bob
measures each logical qubit in the samples by choosing randomly one
of the two bases $\sigma_z^B \equiv \sigma_z^{B_1}\otimes
\sigma_z^{B_2}$ and $\sigma_y^B \equiv \sigma_y^{B_1}\otimes
\sigma_y^{B_2}$. Also he tells Alice which logical qubits are chosen
as samples. Under these two bases, the original entangled state can
be written as
\begin{eqnarray}
\vert \Phi^+_{r} \rangle_{AB}&=& \frac{1}{2}(\vert 0 \rangle_{A}
(\vert 00 \rangle + \vert 11\rangle)_{B_1B_2}\nonumber\\
&&\; + \vert 1 \rangle_{A}
(\vert 01 \rangle - \vert 10 \rangle)_{B_1B_2})\nonumber\\
&=& \frac{1}{\sqrt{2}}(\vert +y \rangle_{A}\vert +y\rangle_{B_1}
\vert -y\rangle_{B_2}\nonumber\\
&& \;\;\;\; + \vert -y \rangle_{A}\vert -y\rangle_{B_1} \vert
+y\rangle_{B_2}).
\end{eqnarray}
That is, the outcomes obtained by Alice and Bob are correlated if
they choose the same bases for their logical qubits. For the message
sequence $S_{BM}$, Bob operates each logical qubit with one of the
four unitary operations
$\{\Theta_{00},\Theta_{01},\Theta_{10},\Theta_{11}\}$ which can be
written as
\begin{eqnarray}
\Theta_{00}&=&\Theta_I=I_1\otimes I_2,\nonumber\\
\Theta_{01}&=&\Theta_z=U_{z1}\otimes U_{z2},\nonumber\\
\Theta_{10}&=&\Theta_x=U_{z1} \otimes U_{x2},\nonumber\\
\Theta_{11}&=&\Theta_y=I_{1} \otimes U_{y2}.
\end{eqnarray}

With these four operations, Bob can manipulate the total state of
the quantum system $AB_1B_2$ and then sent back the two photons
$B_1B_2$ to Alice.  The relation between the initial state, final
state and the operation is shown in Table II. After the operation,
Bob sends the sequence $S_{BM}$ back to Alice.

\begin{center}
\begin{table}[!ht]
\label{table1} \caption{The relation between the initial state
$\vert \psi \rangle_i$, the final state $\vert \psi \rangle_f$ and
the unitary operation.}
\begin{tabular}{c|cccc}\hline
& $\vert \Phi^+_{r} \rangle_f$ & $\vert \Phi^-_{r} \rangle_f$ &
$\vert \Psi^+_{r} \rangle_f$ & $\vert \Psi^-_{r} \rangle_f$
\\\hline $\vert \Phi^+_{r} \rangle_i$ & $\Theta_{00}$ & $\Theta_{01}$ &
$\Theta_{10}$ &
$\Theta_{11}$\\
$\vert \Phi^-_{r} \rangle_i$ & $\Theta_{01}$ & $\Theta_{00}$ &
$\Theta_{11}$ &
$\Theta_{10}$\\
$\vert \Psi^+_{r} \rangle_i$ & $\Theta_{10}$ & $\Theta_{11}$ &
$\Theta_{00}$ &
$\Theta_{01}$\\
$\vert \Psi^-_{r} \rangle_i$ & $\Theta_{11}$ & $\Theta_{10}$ &
$\Theta_{01}$ &
$\Theta_{00}$\\
\hline
\end{tabular}
\end{table}
\end{center}

As these four entangled states in Eq. (\ref{2}) are orthogonal, they
can be discriminated faithfully in principle. With the measurement
results and the initial states, Alice can deduce the unitary
operations performed by Bob. Different from the case with a
collective-dephasing noise, Alice first performs a Hadamard
operation on $B_1 $ and then takes a Bell-state measurement on $A$
and $B_1$ and  a single-photon measurement on the photon $B_2$. The
effect of the Hadamard operation can be written as
\begin{eqnarray} \vert 0 \rangle \rightarrow
\frac{1}{\sqrt{2}}(\vert 0 \rangle +
\vert 1 \rangle),\nonumber\\
\vert 1 \rangle \rightarrow \frac{1}{\sqrt{2}}(\vert 0 \rangle
-\vert 1 \rangle).
\end{eqnarray}
Under the Hadamard operation, the four states in Eq.(\ref{2}) are
transformed into
\begin{eqnarray}
\vert \Phi_{r}^\pm \rangle_{AB_1B_2}\rightarrow \frac{1}{\sqrt{2}}
(\vert \phi^\pm \rangle_{AB_1}\otimes \vert + \rangle_{B_2}-\vert
\psi^\mp \rangle_{AB_1}
\otimes\vert - \rangle_{B_2}),\nonumber\\
\vert \Psi_{r}^\pm \rangle_{AB_1B_2}\rightarrow
\frac{1}{\sqrt{2}}(\vert \psi^\pm \rangle_{AB_1} \otimes \vert +
\rangle_{B_2} - \vert \phi^\mp \rangle_{AB_1} \otimes \vert -
\rangle_{B_2}).\nonumber\\
\end{eqnarray}
The combinations of a  Bell-state measurement and an $X$ basis
single-photon measurement are different for these four states. In
this way, Alice can read out the information about the operations
done by Bob and they can share two bits of information by means of
transmitting two physical qubits in a collective-rotation noise.

As for the security of this quantum cryptography protocol, it can be
made as the same as the case with a collective-dephasing noise.

\section{discussion and summary}

In quantum communication with a noisy channel, the two parties need
introduce some other physical qubits \cite{qkd,lixhqkd}  to prevent
a logical qubit from noise or decrease the success probability of
faithful transmission of qubits \cite{er,dfs1}. When quantum dense
coding is used for quantum cryptography in a noisy channel,
three-photon quantum systems are the optimal ones as the quantum
dense coding based on the measurements on wavepackets is not easy to
be implemented at present. If these two quantum dense coding schemes
are not used for secure quantum communication, they are not optimal
as the effect of the two kinds of collective noise can be avoid with
two orthogonal single-photon states and each qubit can carry one bit
of information. However, only two orthogonal single-photon states
cannot ensure the security of quantum cryptography.

If the losses in the noisy channel is low, our quantum cryptography
schemes can be used for transmitting secret message directly, same
as quantum secure direct communication schemes
\cite{twostep,high,QOTP,yan,gaot,zhangzj}. In this time, Alice and
Bob should first ensure the security of the first transmission,
i.e., that for the sequence $S_B$, and then Bob encodes his secret
message on the sequence $S_{BM}$ with the four unitary operations,
same as Ref. \cite{high}.

In summary, we have presented two fault-tolerant quantum key
distribution protocols against two different kinds of collective
noise, following some ideas in  quantum dense coding. Each logical
qubit, which is invariant when it is transmitted in a
collective-noise channel, is made up of two physical qubits. The
information is encoded on logical qubits with four unitary
operations, which will not destroy the antinoise feather of the
quantum systems. Alice can read out Bob's message with a Bell-state
analysis and a single-photon measurement on each three-qubit quantum
system, not three-qubit joint measurements (i.e., GHZ-state
measurments). Two bits of information are exchanged faithfully and
securely by transmitting two physical qubits through a noisy
channel. When the losses in the noisy channel is low, our quantum
cryptography schemes can, in principle, be used to transmit a secret
message directly.

\section*{ACKNOWLEDGEMENTS}

This work is supported by the National Natural Science Foundation
of China under Grant No. 10604008, A Foundation for the Author of
National Excellent Doctoral Dissertation of P. R. China under
Grant No. 200723, and  Beijing Natural Science Foundation under
Grant No. 1082008.


\begin{thebibliography}{99}
\bibitem{dc} C. H. Bennett and S. J. Wiesner, \emph{Phys. Rev. Lett.} \textbf{69} (1992) 2881.



\bibitem{high} C. Wang, F. G. Deng, Y. S. Li, X. S. Liu and G. L.
Long, \emph{Phys. Rev. A}  \textbf{71}  (2005)  044305.


\bibitem{caipra} Q. Y. Cai and B. W. Li, \emph{Phys. Rev. A}  \textbf{69}  (2004)  054301.

\bibitem{dce} K. Mattle, H. Weinfurter, P. G. Kwiat and  A.
Zeilinger, \emph{Phys. Rev. Lett.} \textbf{76} (1996)   4656.

\bibitem{continuous} S. L. Braunstein and H. J. Kimble, \emph{Phys. Rev. A}
\textbf{61} (2000) 042302.

\bibitem{nuclear} X. Fang, X. Zhu, M. Feng, X. Mao and F. Du, \emph{Phys. Rev. A}
\textbf{61} (2000)  022307.

\bibitem{atomic} T. Schaetz, M. D. Barrett, D. Leibfried, J. Chiaverini, J. Britton, W. M. Itano, J. D. Jost, C.
Langer and
 D. J. Wineland, \emph{Phys. Rev. Lett.}  \textbf{93} (2004)
 040505.


\bibitem{multi} X. S. Liu, G. L. Long, D. M. Tong and  L. Feng, \emph{Phys. Rev. A}
\textbf{65} (2002)
 022304.

\bibitem{hyper} J. T. Barreiro, T. C. Wei and  P. G. Kwiat, \emph{Nature
Physics}  \textbf{4}   (2008)  282.

\bibitem{kerr} K. Nemoto and W. J. Munro, \emph{Phys. Rev. Lett.}
\textbf{93} (2004)
 250502.

\bibitem{momentum} P. G. Kwiat and H. Weinfurter, \emph{Phys. Lett. A}
\textbf{58} (1998)
  R2623.

\bibitem{complete} C. Schuck, G. Huber, C. Kurtsiefer  and H.
Weinfurter, \emph{Phys. Rev. Lett.}  \textbf{96}  (2006) 190501.

\bibitem{collective} P. Zanardi and M. Rasetti, \emph{Phys. Rev. Lett.}  \textbf{79}  (1997)  3306.

\bibitem{ep} C. H. Bennett, G. Brassard, S. Popescu, B. Schumacher, J. A. Smolin and  W. K.
Wootters,  \emph{Phys. Rev. Lett.}  \textbf{76}  (1996)  722.

\bibitem{qecc} M. A. Nielsen, I. L. Chuang, \emph{Quantum Computation
and Quantum Information} (Cambridge University Press, Cambridge,
England, 2000).



\bibitem{er} X. H. Li, F. G. Deng and   H. Y. Zhou, Appl. Phys. Lett.
\textbf{91} (2007)
 144101.

\bibitem{dfs1} Z. D. Walton, A. F. Abouraddy, A. V. Sergienko, B. E. A. Saleh and  M. C.
Teich, \emph{Phys. Rev. Lett.}  \textbf{91}  (2003)  087901.

\bibitem{qkd} J. C. Boileau, D. Gottesman, R. Laflamme, D. Poulin and  R. W. Spekkens,  Phys. Rev. Lett.  92 (2004) 017901.






\bibitem{bb84} C. H. Bennett and G. Brassard, in: Proceedings of IEEE
International Conference on Computers, Systems and Signal
Processing, (Bangalore, India, IEEE, New York, 1984),  p 175.

\bibitem{b92} C. H. Bennett, \emph{Phys. Rev. Lett.} \textbf{68} (1992)  3121.


\bibitem{Ekert91} A. K. Ekert, \emph{Phys. Rev. Lett.}  \textbf{67}   (1991)  661.

\bibitem{BBM92} C. H. Bennett, G. Brassard  and N. D. Mermin, \emph{Phys. Rev. Lett.}
 \textbf{68}   (1992)  557.

\bibitem{rmp} N. Gisin, G. Ribordy, W. Tittel and H.
Zbinden, \emph{Rev. Mod. Phys.}  \textbf{74}   (2002) 145.

\bibitem{longqkd} G. L. Long and X. S. Liu, \emph{Phys. Rev. A}  \textbf{65}
(2002)  032302.

\bibitem{CORE} F. G. Deng and G. L. Long, \emph{Phys. Rev. A}
\textbf{68}   (2003)  042315.

\bibitem{BidQKD} F. G. Deng and G. L. Long, \emph{Phys. Rev. A}
 \textbf{70}  (2004)  012311.

\bibitem{lixhqkd} X. H. Li, F. G. Deng and  H. Y. Zhou,  \emph{Phys. Rev.  A}
\textbf{78} (2008) 022321.




\bibitem{twostep} F. G. Deng, G. L. Long and X. S. Liu, \emph{Phys. Rev. A}
 \textbf{68}  (2003)  042317.

\bibitem{spy} X. H. Li, F. G. Deng and  H. Y. Zhou, \emph{Phys. Rew. A}
\textbf{74}  (2006) 054302.

\bibitem{prove1} H. Inamori, L. Rallan, and  V. Vedral, \emph{J. Phys. A} \textbf{34}
(2001)  6913.

\bibitem{prove2} E. Waks, A. Zeevi and Y. Yamamoto, \emph{Phys. Rev. A}
\textbf{65} (2002) 052310.

\bibitem{QOTP} F. G. Deng and G. L. Long, \emph{Phys. Rev. A} \textbf{69}
(2004)  052319.


\bibitem{yan} F. L.  Yan and X. Zhang,
\emph{Euro. Phys. J. B } \textbf{41} (2004)  75.

\bibitem{gaot} T. Gao, F. L. Yan and Z. X. Wang, \emph{J. Phys. A} \textbf{38}
(2005) 5761.

\bibitem{zhangzj} Z. X. Man, Z. J. Zhang and Y. Li,
\emph{Chin. Phys. Lett.} \textbf{22} (2005) 18.







\end{thebibliography}
\end{document}